\documentclass[twocolumn,floatfix,showpacs,superscriptaddress]{revtex4}
\pdfoutput=1
\usepackage{graphicx}
\usepackage{amsmath}
\usepackage{bm}
\begin{document}

\title{Switching of electrical current by spin precession in the first Landau level of an inverted-gap semiconductor}

\author{A. R. Akhmerov}
\affiliation{Instituut-Lorentz, Leiden University, P.O. Box 9506, 2300 RA Leiden, The Netherlands}
\author{C. W. Groth}
\affiliation{Instituut-Lorentz, Leiden University, P.O. Box 9506, 2300 RA Leiden, The Netherlands}
\author{J. Tworzyd{\l}o}
\affiliation{Institute of Theoretical Physics, Warsaw University, Ho\.{z}a 69, 00--681 Warsaw, Poland}
\author{C. W. J. Beenakker}
\affiliation{Instituut-Lorentz, Leiden University, P.O. Box 9506, 2300 RA Leiden, The Netherlands}

\date{June 2009}
\begin{abstract}
We show how the quantum Hall effect in an inverted-gap semiconductor (with electron- and hole-like states at the conduction- and valence-band edges interchanged) can be used to inject, precess, and detect the electron spin along a one-dimensional pathway. The restriction of the electron motion to a single spatial dimension ensures that all electrons experience the same amount of precession in a parallel magnetic field, so that the full electrical current can be switched on and off. As an example, we calculate the magnetoconductance of a \textit{p-n} interface in a HgTe quantum well and show how it can be used to measure the spin precession due to bulk inversion asymmetry.
\end{abstract}
\maketitle

\section{Introduction}

A central goal of spin-transport electronics (or spintronics) is the ability to switch current between spin-selective electrodes by means of spin precession \cite{Zut04}. In the original Datta-Das proposal for such a spin-based transistor \cite{Dat90}, the current which is switched carries both spin and charge. It has proven difficult to separate the effects of spin precession from purely orbital effects (deflection of electron trajectories), so most succesful implementations use a nonlocal geometry \cite{Joh85} to modulate the spin current at zero charge current \cite{Jed01,Lou07,Tom07}. Even in the absence of an orbital effect, the fact that different electrons (moving along different trajectories) experience different amounts of spin precession prevents a complete switching of the current from one electrode to the other.  

If the electron motion could somehow be confined to a single spatial dimension, it would be easier to isolate spin effects from orbital effects and to ensure that all electron spins precess by the same amount. Complete switching of the current would then be possible, limited only by spin relaxation processes. Edge state transport in the quantum Hall effect is one-dimensional and spin selective (in sufficiently strong perpendicular magnetic fields $B_{\perp}$), but spin precession plays no role in the traditional experiments on a two-dimensional electron gas \cite{Bee91}. In this paper we show how the quantum Hall effect in an inverted-gap semiconductor offers the unique possibility to perform a one-dimensional spin precession experiment.

\begin{figure}[tb]
\centerline{\includegraphics[width=0.8\linewidth]{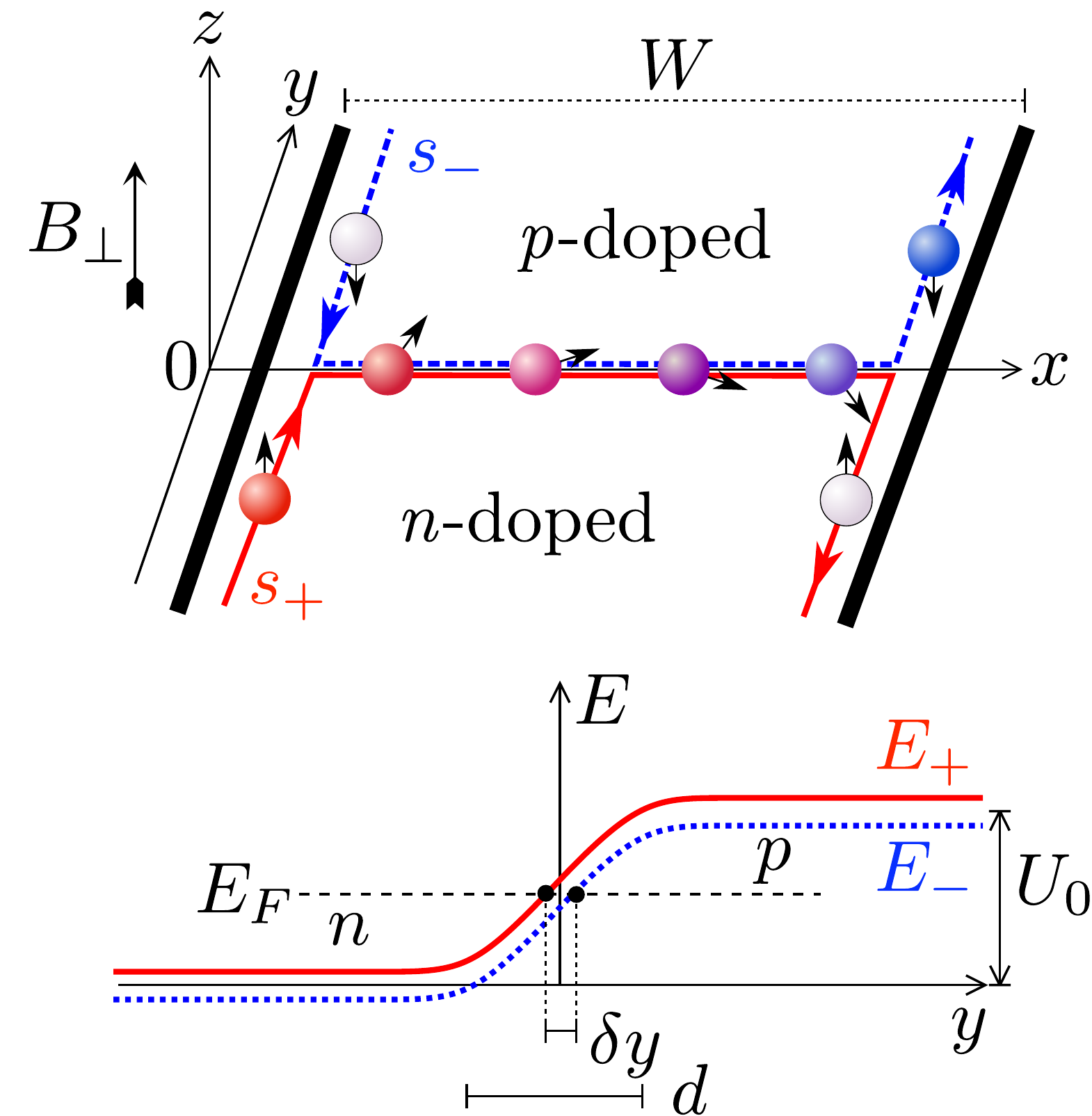}}
\caption{\label{fig_geometry}
Top panel: Schematic illustration of the one-dimensional pathway along which the electron spin is injected, precessed, and detected (filled circles: occupied states; open circles: empty states). Bottom panel: Potential profile of the \textit{p-n} junction, shown for $B_{\perp}>B_{c}$ (for $B_{\perp}<B_{c}$ the labels $E_{+}$ and $E_{-}$ should be interchanged).}
\end{figure}

The key idea is to combine the spin-selectivity of edge states with free precession along a \textit{p-n} interface. The geometry, shown in Fig.\ \ref{fig_geometry}, has been studied in graphene \cite{Wil07,Aba07,Ozy07,Two07} --- but there spin is only weakly coupled to the orbit and plays a minor role \cite{Kan05,Aba06}. The strong spin-orbit coupling in inverted-gap semiconductors splits the first Landau level into a pair of levels $E_{\pm}$ of opposite magnetic moment \cite{Kon08,Sch09}. 
One level $E_{+}$ (say, with spin up) has electron-like character and produces edge states in the conduction band. The other level $E_{-}$ (with spin down) has hole-like character and produces edge states in the valence band. The edge states from $E_{+}$ and $E_{-}$ have opposite chirality, meaning that one circulates clockwise along the edge while the other circulates counter-clockwise. These spin-selective, chiral edge states provide the spin injection at $x=0$ and detection at $x=W$. 

For the spin precession we need to combine states from $E_{+}$ and $E_{-}$. This is achieved by means of a gate electrode, which creates a smooth potential step (height $U_{0}$, width $d$) centered at $y=0$, such that the Fermi level lies in the conduction band for $y<0$ (\textit{n}-doped region) and in the valence band for $y>0$ (\textit{p}-doped region). At the \textit{p-n} interface states from the first Landau levels $E_{+}$ and $E_{-}$ overlap at the Fermi energy $E_{F}^{}$, to form a spin-degenerate one-dimensional state. Spin precession can be realized externally by a parallel magnetic field $\bm{B}_{\parallel}$  (in the $x-y$ plane) or internally by bulk or structure inversion asymmetry \cite{Kon08}.

Good overlap at $E_{F}^{}$ of the states from $E_{+}$ and $E_{-}$ is crucial for effective spin precession. The requirement is that the spatial separation $\delta y\simeq|E_{+}-E_{-}|d/U_{0}$ of the states should be small compared to the magnetic length $l_{m}=(\hbar/eB_{\perp})^{1/2}$ (which sets their spatial extent). This is where the inverted gap comes in, as we now explain. 

Inversion of the gap means that the first Landau level in the conduction band goes down in energy with increasing magnetic field (because it has hole-like character), while the first Landau level in the valence band goes up in energy (because it has electron-like character). As a consequence, the gap $|E_{+}-E_{-}|$ has a minimal value $E_{c}$ much less than the cyclotron energy $\hbar\omega_{c}$ at a crossover magnetic field $B_{c}$. Indeed, $E_{c}=0$ in the absence of inversion asymmetry \cite{Kon08}. Good overlap can therefore be reached in an inverted-gap semiconductor, simply by tuning the magnetic field. In a normal (non-inverted) semiconductor, such as GaAs, the cyclotron energy difference between $E_{+}$ and $E_{-}$ effectively prevents the overlap of Landau levels from conduction and valence bands.  

In the following two sections, we first present a general, model independent analysis and then specialize to the case of a HgTe quantum well (where we test the analytical theory by computer simulation).

\section{General theory}

We introduce a one-dimensional coordinate $s_{\pm}$ along the $E_{\pm}$ edge states, increasing in the direction of the chirality (see Fig.\ \ref{fig_geometry}). The wave amplitudes $\psi_{\pm}(s_{\pm})$ of these two states can be combined into the spinor
$\Psi=(\psi_{+},\psi_{-})$. Far from the \textit{p-n} interface, $\psi_{+}$ and $\psi_{-}$ evolve independently with Hamiltonian
\begin{equation}
H_{0}=\begin{pmatrix}
H_{+}&0\\
0&H_{-}
\end{pmatrix},\;\;
H_{\pm}=v_{\pm}\left(-i\hbar \frac{\partial}{\partial s_{\pm}}-p_{F}^{\pm}\right).\label{Hpmdef}
\end{equation}
This is the generic linearized Hamiltonian of a chiral mode, with group velocity $v_{\pm}\equiv v(s_{\pm})$ and Fermi momentum $p_{F}^{\pm}\equiv p_{F}^{}(s_{\pm})$. Near the \textit{p-n} interface the spin-up and spin-down states are coupled by the generic precession Hamiltonian,
\begin{equation}
 H_{\rm prec}=\begin{pmatrix}
0&{\cal M}^*\\
{\cal M}&0
\end{pmatrix},\label{Hprecdef}
\end{equation}
with a matrix element ${\cal M}$ to be specified later.

We seek the transfer matrix $T$, defined by
\begin{equation}
\Psi(s^{f}_{+},s^{f}_{-})= T\Psi(s^{i}_{+},s^{i}_{-}).
\end{equation}
We take for $\Psi$ a solution of the Schr\"{o}dinger equation,
\begin{equation}
(H_0+H_{\rm prec})\Psi=0,\label{Schrodinger}
\end{equation}
at zero excitation energy (appropriate for electrical conduction in linear response). The initial and final points $s_{\pm}^{i}$ and $s_{\pm}^{f}$ are taken away from the \textit{p-n} interface. The unitary scattering matrix $S$ (relating incident and outgoing current amplitudes) is related to $T$ by a similarity transformation,
\begin{equation}
S=\begin{pmatrix}
v_{+}^{f}&0\\
0&v_{-}^{f}
\end{pmatrix}^{1/2}T\begin{pmatrix}
v_{+}^{i}&0\\
0&v_{-}^{i}
\end{pmatrix}^{-1/2}.\label{Sdef}
\end{equation}
The two-terminal linear-response conductance $G$ of the \textit{p-n} junction is given by the Landauer formula,
\begin{equation}
G=\frac{e^{2}}{h}|S_{21}|^{2}.\label{Landauer}
\end{equation}

The transition matrix element ${\cal M}(s_{+},s_{-})$ between $\psi_{+}(s_{+})$ and $\psi_{-}(s_{-})$ vanishes if the separation $|s_{+}-s_{-}|$ of the two states is large compared to the magnetic length $l_{m}$. We assume that $B_{\perp}$ is sufficiently close to $B_{c}$ that $|s_{+}-s_{-}|<l_{m}$ at the \textit{p-n} interface $y=0$, $0<x<W$, where we may take ${\cal M}={\rm constant}$ (independent of $x$). At the two edges $x=0$ and $x=W$ we set ${\cal M}=0$, neglecting the crossover region within $l_{m}$ of $(0,0)$ and $(W,0)$. (The precession angle there will be small compared to unity for $l_{m}\ll \hbar v_{\pm}/|{\cal M}|$.)

In this ``abrupt approximation'' we may identify the initial and final coordinates $s_{\pm}^{i}$ and $s_{\pm}^{f}$ with the points $(0,0)$ and $(W,0)$, at the two ends of the \textit{p-n} interface. Integration of the Schr\"{o}dinger equation \eqref{Schrodinger} along the \textit{p-n} interface gives the transfer matrix, and application of Eq. \eqref{Sdef} then gives the scattering matrix
\begin{equation}
S=\exp\left[-i \frac{W}{\hbar}
\begin{pmatrix}
 p_{F}^{+} & {\cal M}^*/\sqrt{v_{+}v_{-}}\\
{\cal M}/\sqrt{v_{+}v_{-}} & p_{F}^{-}
\end{pmatrix}
\right].\label{Sresult}
\end{equation}
(We have assumed that $v_{\pm}$ and $p_{F}^{\pm}$, as well as ${\cal M}$, do not vary along the \textit{p-n} interface, so we may omit the labels $i,f$.) One verifies that $S$ is unitary, as it should be.

Evaluation of the matrix exponent in Eq.\ \eqref{Sresult} and substitution into Eq.\ \eqref{Landauer} gives the conductance,
\begin{equation}
G=\frac{e^{2}}{h}\sin^{2}\left(\frac{|\bm{p}_{\rm eff}|W}{\hbar}\right)\sin^{2}\alpha.\label{Gresult}
\end{equation}
The effective precession momentum
\begin{equation}
\bm{p}_{\rm eff}=\left(\frac{\textrm{Re}\,{\cal M}}{\bar{v}},\frac{\textrm{Im}\,{\cal M}}{\bar{v}},\frac{\delta p_{F}}{2}\right)\label{meffdef}
\end{equation}
(with $\delta p_{F}=p_{F}^{+}-p_{F}^{-}$ and $\bar{v}=\sqrt{v_{+}v_{-}}$)
makes an angle $\alpha$ with the $z$-axis. This is the final result of our general analysis.

\section{Application to a ${\rm\bf HgTe}$ quantum well}

We now turn to a specific inverted-gap semiconductor, a quantum well consisting of a $7\,{\rm nm}$ layer of HgTe sandwiched symmetrically between ${\rm Hg}_{0.3}{\rm Cd}_{0.7}{\rm Te}$ \cite{Kon07}. The properties of this socalled topological insulator have been reviewed recently \cite{Kon08}. The low-energy excitations are described by a four-orbital tight-binding Hamiltonian \cite{Ber06,Fu07},
\begin{equation}
H=\sum_{n}c^{\dagger}_{n}{\cal E}_{n}c^{\vphantom{\dagger}}_{n}-\!\!\!\!\sum_{n,m\;{\rm (nearest\; neighb.)}}c_{n}^{\dagger}{\cal T}_{nm}c^{\vphantom{\dagger}}_{m}.\label{Htb}
\end{equation}
Each site $n$ on a square lattice (lattice constant $a=4\,{\rm nm}$) has four states $|s,\pm\rangle$, $|p_{x}\pm ip_{y},\pm\rangle$ --- two electron-like $s$-orbitals and two hole-like $p$-orbitals of opposite spin $\sigma=\pm$. Annihilation operators $c_{n,\tau\sigma}$ for these four states (with $\tau\in\{s,p\}$) are collected in a vector
\[
c_{n}=(c_{n,s+},c_{n,p+},c_{n,s-},c_{n,p-}). 
\]
States on the same site are coupled by the $4\times 4$ potential matrix ${\cal E}_{n}$ and states on adjacent sites by the $4\times 4$ hopping matrix ${\cal T}_{nm}$.

In zero magnetic field and without inversion asymmetry $H$ decouples into a spin-up block $H_{+}$ and a spin-down block $H_{-}$, defined in terms of the $2\times 2$ matrices
\begin{align}
&{\cal E}_{n}^{+}={\cal E}_{n}^{-}={\rm diag}\,(\varepsilon_{s}-U_n,\varepsilon_{p}-U_n),\label{Epmdef}\\
&{\cal T}_{nm}^{+}=\left({\cal T}_{nm}^{-}\right)^{\ast}=\begin{pmatrix}
t_{ss}&t_{sp}e^{i\theta_{nm}}\\
t_{sp}e^{-i\theta_{mn}}&-t_{pp}
\end{pmatrix}.\label{Tpmdef}
\end{align}
Here $U_n$ is the electrostatic potential and $\theta_{nm}$ is the angle between the vector $\bm{r}_{n}-\bm{r}_{m}$ and the positive $x$-axis (so $\theta_{mn}=\pi-\theta_{nm}$). The orbital effect of a  perpendicular magnetic field $B_{\perp}$ is introduced into the hopping matrix elements by means of the Peierls substitution
\[
{\cal T}_{nm}\mapsto{\cal T}_{nm}\exp[i(eB_{\perp}/\hbar)(y_{n}-y_{m})x_{n}]. 
\]
This breaks the degeneracy of the spin-up and spin-down energy levels, but it does not couple them. 

Spin-up and spin-down states are coupled by the Zeeman effect from a parallel magnetic field (with gyromagnetic factor $g_{\parallel}$) and by spin-orbit interaction without inversion symmetry (parameterized by a vector $\bm{\Delta}$). In first-order perturbation theory, the correction $\delta{\cal E}$ to the on-site potential has the form \cite{Kon08}
\begin{align}
\delta{\cal E}={}&(\bm{\Delta}\cdot\bm{\sigma})\otimes\tau_{y}+\tfrac{1}{2}\mu_{B}g_{\parallel}(\bm{B}_{\parallel}\cdot\bm{\sigma})\otimes(\tau_{0}+\tau_z)\nonumber\\
&+\mu_{B}B_{\perp}\sigma_{z}\otimes(\bar{g}_{\perp}\tau_{0}+\delta g_{\perp}\tau_{z}).\label{deltaE}
\end{align}
The Pauli matrices $\bm{\sigma}=(\sigma_{x},\sigma_{y},\sigma_{z})$ act on the spin-up and spin-down blocks, while the Pauli matrices $\tau_{y},\tau_{z}$ and the unit matrix $\tau_{0}$ act on the orbital degree of freedom $s,p$ within each block.

The parameters of the tight-binding model for a $7\,{\rm nm}$ thick ${\rm HgTe}/{\rm Hg}_{0.3}{\rm Cd}_{0.7}{\rm Te}$ quantum well (grown in the $(001)$ direction) are as follows \cite{Kon08}: $t_{ss}=74.9\,{\rm meV}$, $t_{pp}=10.9\,{\rm meV}$, $t_{sp}=45.6\,{\rm meV}$, $\varepsilon_s=289.5\,{\rm meV}$, $\varepsilon_p=-33.5\,{\rm meV}$, $\bar{g}_{\perp}=10.75$, $\delta g_{\perp}=11.96$, $g_\parallel=-20.5$, $\bm{\Delta}=(0,1.6\,{\rm meV},0)$.

\begin{figure}[tb]
\includegraphics[width=1\linewidth]{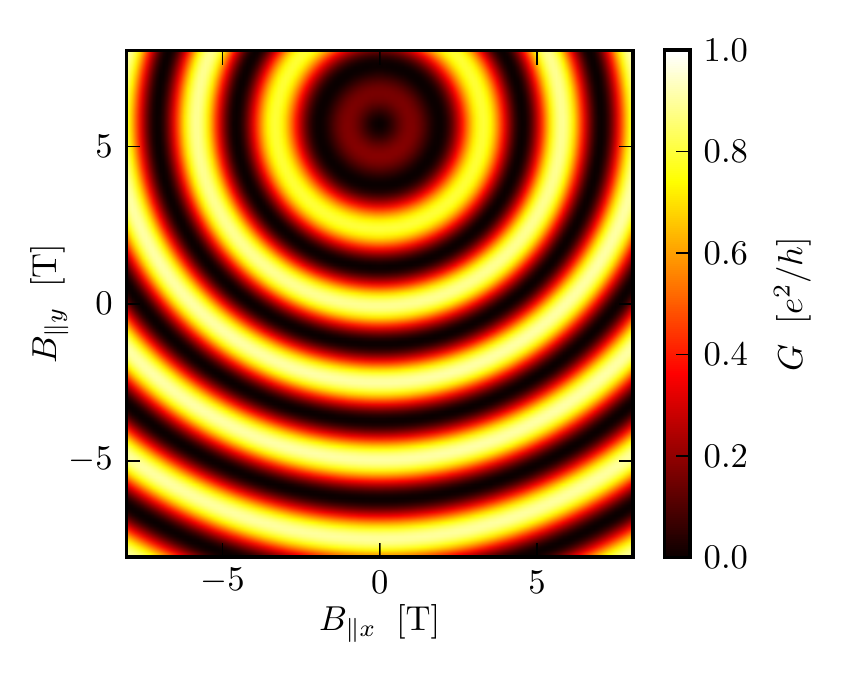}
\caption{\label{fig:bullseye} Dependence of the conductance of the HgTe quantum well on the parallel magnetic field $\bm{B}_{\parallel}$, calculated from the tight-binding model for $B_{\perp}=B_{c}=6.09\,{\rm T}$.}
\end{figure}

\begin{figure}[tb]
\includegraphics[width=0.8\linewidth]{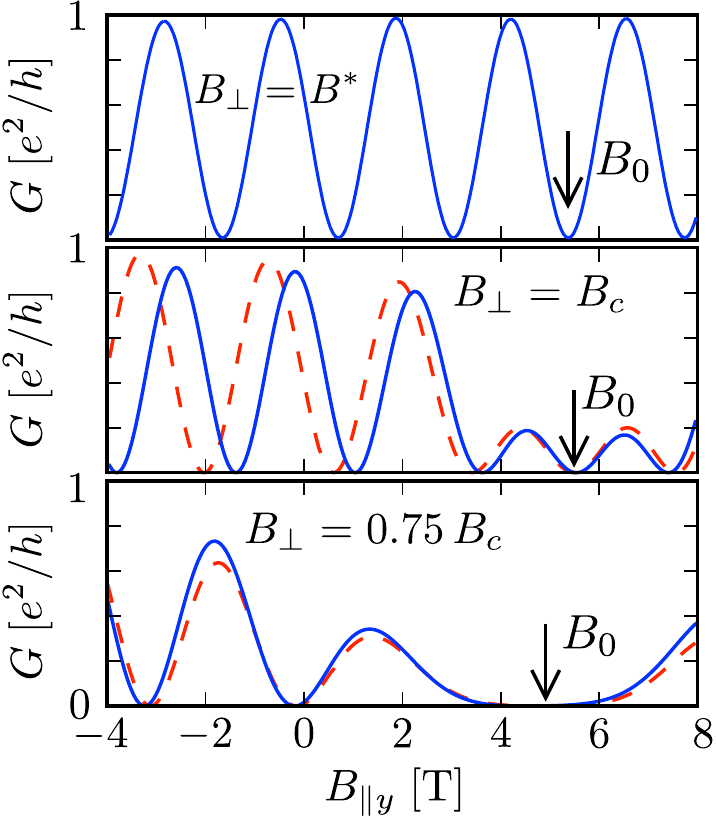}
\caption{\label{fig:scanline} Dependence of the conductance on $B_{\parallel y}$ for $B_{\parallel x}=0$, at three values of the perpendicular magnetic field. The solid curves are calculated numerically from the tight-binding model, the dashed curves are the analytical prediction \eqref{Gbullseye}. The arrow indicates the value of $B_{0}$ from Eq.\ \eqref{Bpar0}. (Only the numerical curve is shown in the upper panel, because the analytical curve is nearly indistinguishable from it.)}
\end{figure}

The quantum well is symmetric, so only bulk inversion asymmetry contributes to $\bm{\Delta}$. The \textit{p-n} junction is defined by the potential profile
\begin{equation}
 U(x,y)=\tfrac{1}{2}U_0[1+\tanh(4y/d)],\;\;0<x<W,\label{Udef}
\end{equation}
with $U_0=32\,{\rm meV}$, $d=12\,{\rm nm}$, and $W=0.8\,\mu{\rm m}$. We fix the Fermi level at $E_F=25\,{\rm meV}$, so that it lies in the conduction band for $y<0$ and in the valence band for $y>0$. (We have checked that none of the results are sensitive to the choice of potential profile or parameter values.) The scattering matrix of the \textit{p-n} junction is calculated with the recursive Green function technique, using the ``knitting'' algorithm of Ref.\ \cite{Kaz08}. Results for $G$ as a function of $\bm{B}_{\parallel}$ are shown in Figs.\ \ref{fig:bullseye} and \ref{fig:scanline}.

The dependence of the conductance on the parallel magnetic field $\bm{B}_{\parallel}$ shows a striking ``bullseye'' pattern, which can be understood as follows. To first order in $\bm{B}_{\parallel}$, the edge state parameters $v_{\pm}$ and $p_{F}^{\pm}$ are constant, while the precession matrix element
\begin{equation}
{\cal M}={\Delta}_{\rm eff}+\mu_{B}g_{\rm eff}(B_{\parallel x}+iB_{\parallel y})
\end{equation}
varies linearly. Substitution into Eqs.\ \eqref{Gresult} and \eqref{meffdef} gives a circularly symmetric dependence of $G$ on $\bm{B}_{\parallel}$,
\begin{align} 
&G=\frac{e^2}{h}\left(1+\frac{(\bar{v}\delta p_F)^2}{4|\mu_{B}g_\textrm{eff}|^2|\bm{B}_{\parallel}-\bm{B}_{0}|^2}\right)^{-1}\nonumber\\
&\;\;\times \sin^2\left[\frac{W}{\hbar\bar{v}} \sqrt{|\mu_{B}g_\textrm{eff}|^2|\bm{B}_{\parallel}-\bm{B}_{0}|^2+\tfrac{1}{4}(\bar{v}\delta p_F)^2}\right]\label{Gbullseye},\\
&\bm{B}_{0}=\mu_{B}^{-1}\biglb(\textrm{Re}[\Delta_{\rm eff}/g_{\rm eff}],\textrm{Im}
[\Delta_{\rm eff}/g_{\rm eff}],0\bigrb).\label{Bpar0}
\end{align}
The parallel magnetic field $\bm{B}_{0}$ corresponds to the center of the bullseye, at which the coupling between the $\pm$ edge states along the \textit{p-n} interface by bulk inversion asymmetry is cancelled by the Zeeman effect.

The Fermi momentum mismatch $\delta p_{F}$ vanishes at a perpendicular magnetic field $B^{\ast}$ close to, but not equal to, $B_{c}$. Then the magnetoconductance oscillations are purely sinusoidal,
\begin{equation}
G=\frac{e^{2}}{h}\sin[(W/\hbar\bar{v})\mu_{B}g_{\textrm{eff}}|\bm{B}_{\parallel}-\bm{B}_{0}|].\label{Gsine}
\end{equation}

For a quantitative comparison between numerics and analytics, we extract the parameters $v_{\pm}$ and $p_{F}^{\pm}$ from the dispersion relation of the edge states $\psi_{\pm}$ along an infinitely long \textit{p-n} interface (calculated for uncoupled blocks $H_{\pm}$). The overlap of $\psi_{+}$ and $\psi_{-}$ determines the coefficients 
\begin{align}
&\Delta_{\rm eff}=(\Delta_x+i\Delta_y)\langle\psi_{-}|\tau_{y}|\psi_{+}\rangle,\label{Deltaeff}\\
&g_{\rm eff}=\tfrac{1}{2}g_\parallel \langle\psi_{-}|\tau_{0}+\tau_{z}|\psi_{+}\rangle.\label{geff}
\end{align}

For $B_{\perp}=B_{c}=6.09\,{\rm T}$ we find $\bar{v}\delta p_{F}=0.86\,{\rm meV}$, $\hbar\bar{v}/W=0.23\,{\rm meV}$, $\Delta_{\rm eff}=-1.59\,{\rm meV}$, $g_{\rm eff}=-4.99$. The Fermi momentum mismatch $\delta p_{F}$ vanishes for $B_{\perp}=B^{\ast}=5.77\,{\rm T}$. Substitution of the parameters into Eq.\ \eqref{Gbullseye} gives the dashed curves in Fig.\ \ref{fig:scanline}, in reasonable agreement with the numerical results from the tight-binding model (solid curves). In particular, the value of $B_{0}$ extracted from the numerics is within a few percent of the analytical prediction \eqref{Bpar0}.

\begin{figure}[tb]
\includegraphics[width=0.8\linewidth]{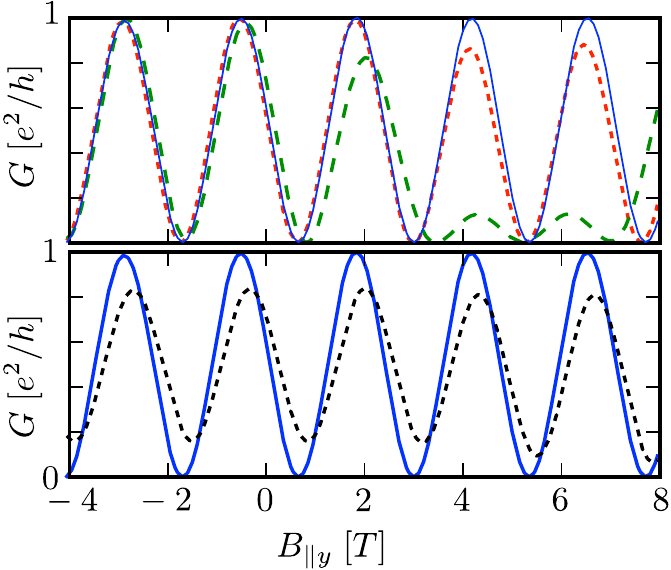}
\caption{\label{fig:temp_disorder} The solid blue curve in both panels is the same as in Fig.\ \ref{fig:scanline}, top panel, calculated for $B_{\perp}=B^{\ast}$ from the tight-binding model at zero temperature without any disorder. The dotted black curve in the lower panel shows the effect of raising the temperature to $30\,{\rm K}\approx U_0/3k_{B}$. The dotted red curve and dashed green curve in the upper panel show the effect of disorder at zero temperature. The on-site disorder potential is drawn uniformly from the interval $(-\Delta U_{0},\Delta U_{0})$, with, respectively, $\Delta U=U_{0}/4$ and $\Delta U=U_{0}/2$.
}
\end{figure}

Because of the one-dimensionality of the motion along the \textit{p-n} interface, electrostatic disorder and thermal averaging have a relatively small perturbing effect on the conductance oscillations. For disorder potentials $\Delta U$ and thermal energies $k_{B}T$ up to 10\% of $U_{0}$ the perturbation is hardly noticeable (a few percent). As shown in Fig.\ \ref{fig:temp_disorder}, the conductance oscillations remain clearly visible even for $\Delta U$ and $k_{B}T$ comparable to $U_{0}$. In particular, we have found that the center of the bullseye pattern remains within $10\%$ of $B_{0}$ even for $\Delta U$ as large as the \textit{p-n} step height $U_{0}$.

\section{Conclusion}
 
In conclusion, we have proposed a one-dimensional spin precession experiment at a \textit{p-n} junction in an inverted-gap semiconductor. The conductance as a function of parallel magnetic field oscillates in a bullseye pattern, centered at a field $B_{0}$ proportional to the matrix element $\Delta_{\rm eff}$ of the bulk inversion asymmetry. Our numerical and analytical calculations show conductance oscillations of amplitude not far below $e^{2}/h$, robust to disorder and thermal averaging. Realization of the proposed experiment in a HgTe quantum well \cite{Kon08} (or in other inverted-gap semiconductors \cite{Liu08a}) would provide a unique demonstration of full-current switching by spin precession. 

As directions for future research, we envisage potential applications of this technique as a sensitive measurement of the degree of bulk inversion asymmetry, or as a probe of the effects of interactions on spin precession. It might also be possible to eliminate the external magnetic field and realize electrical switching of the current in our setup: The role of the perpendicular magnetic field in producing spin-selective edge states can be taken over by magnetic impurities or a ferromagnetic layer \cite{Liu08b}, while the role of the parallel magnetic field in providing controlled spin precession can be taken over by gate-controlled structural inversion asymmetry.

\acknowledgments

This research was supported by the Dutch Science Foundation NWO/FOM and by an Initial Training Network (NanoCTM) of the European Community . We acknowledge helpful correspondence with C.-X. Liu and X.-L. Qi.


\begin{thebibliography}{99}
\bibitem{Zut04} I. \v{Z}uti\'{c}, J. Fabian, and S. Das Sarma, Rev. Mod. Phys. \textbf{76}, 323 (2004).
\bibitem{Dat90} S. Datta and B. Das, Appl. Phys. Lett. \textbf{56}, 665 (1990).
\bibitem{Joh85} M. Johnson and R. H. Silsbee, Phys. Rev. Lett. \textbf{55}, 1790 (1985). 
\bibitem{Jed01} F. J. Jedema, H. B. Heersche, A. T. Filip, J. J. A. Baselmans, and B. J. van Wees,  Nature \textbf{416}, 713 (2001).
\bibitem{Lou07} X. Lou, C. Adelmann, S. A. Crooker, E. S. Garlid, J. Zhang, K. S. Madhukar Reddy, S. D. Flexner, C. J. Palmstrom, and P. A. Crowell, Nature Phys. \textbf{3}, 197 (2007). 
\bibitem{Tom07} N. Tombros, C. Jozsa, M. Popinciuc, H. T. Jonkman, and B. J. Van Wees, Nature \textbf{448}, 571 (2007). 
\bibitem{Bee91} C. W. J. Beenakker and H. van Houten, Solid State Phys. \textbf{44}, 1 (1991); cond-mat/0412664.
\bibitem{Wil07} J. R. Williams, L. DiCarlo, and C. M. Marcus, Science {\bf 317}, 638 (2007).
\bibitem{Aba07} D. A. Abanin and L. S. Levitov, Science {\bf 317}, 641 (2007).
\bibitem{Ozy07} B. \"{O}zyilmaz, P. Jarillo-Herrero, D. Efetov, D. A. Abanin, L. S. Levitov, and P. Kim, Phys.\ Rev.\ Lett.\ {\bf 99}, 166804 (2007).
\bibitem{Two07} J. Tworzyd{\l}o, I. Snyman, A. R. Akhmerov, and C. W. J. Beenakker, Phys.\ Rev.\ B {\bf 76}, 035411 (2007).
\bibitem{Kan05} C. L. Kane and E. J. Mele, Phys. Rev. Lett. \textbf{95}, 226801 (2005).
\bibitem{Aba06} D. A. Abanin, P. A. Lee, and L. S. Levitov, Phys. Rev. Lett. \textbf{96}, 176803 (2006).
\bibitem{Kon08} M. K\"{o}nig, H. Buhmann, L. Molenkamp, T. Hughes, C.-X. Liu, X.-L. Qi, and S.-C. Zhang, J. Phys. Soc. Japan \textbf{77}, 031007 (2008).
\bibitem{Sch09} M. J. Schmidt, E. G. Novik, M. Kindermann, and B. Trauzettel, Phys. Rev. B \textbf{79}, 241306(R) (2009).
\bibitem{Kon07} M. K\"{o}nig, S. Wiedmann, C. Br\"{u}ne, A. Roth, H. Buhmann, L. Molenkamp, X.-L. Qi, and S.-C. Zhang, Science \textbf{318}, 766 (2007).
\bibitem{Ber06} B. A. Bernevig, T. L. Hughes, and S.-C. Zhang, Science \textbf{314}, 1757 (2006).
\bibitem{Fu07} L. Fu and C. L. Kane, Phys. Rev. B \textbf{76}, 045302 (2007).
\bibitem{Kaz08} K. Kazymyrenko and X. Waintal, Phys. Rev. B \textbf{77}, 115119 (2008). The computer code for the knitting algorithm was kindly provided to us by Dr.\ Waintal.
\bibitem{Liu08a} C.-X. Liu, T. Hughes, X.-L. Qi, K. Wang, and S.-C. Zhang, Phys. Rev. Lett. \textbf{100}, 236601 (2008).
\bibitem{Liu08b} C.-X. Liu, X.-L. Qi, X. Dai, Z. Fang, and S.-C. Zhang, Phys. Rev. Lett. \textbf{101}, 146802 (2008).
\end{thebibliography}
\end{document}